\documentclass{appolb}
\usepackage{epsfig}

\usepackage{bm}%
\usepackage[colorlinks=true,linkcolor=red]{hyperref}
\usepackage{amsmath}
\usepackage{amsfonts}
\usepackage{amssymb}
\usepackage{cancel}
\usepackage{eucal}
\usepackage{hyperref}
\usepackage{verbatim}

\newcommand{\newc}{\newcommand}
\newc{\beq}{\begin{equation}}
\newc{\eeq}{\end{equation}}
\newc{\kt}{\rangle}
\newc{\bra}{\langle}
\newc{\beqa}{\begin{eqnarray}}
\newc{\eeqa}{\end{eqnarray}}
\newc{\pr}{\prime}
\newc{\longra}{\longrightarrow}
\newc{\ot}{\otimes}
\newc{\rarrow}{\rightarrow}
\newc{\h}{\hat}
\newc{\bom}{\boldmath}
\newc{\btd}{\bigtriangledown}
\newc{\al}{\alpha}
\newc{\be}{\beta}
\newc{\ld}{\lambda}
\newc{\ldmin}{\lambda_{\rm min}}
\newc{\sg}{\sigma}
\newc{\p}{\psi}
\newc{\eps}{\epsilon}
\newc{\om}{\omega}
\newc{\mb}{\mbox}
\newc{\tm}{\times}
\newc{\hu}{\hat{u}}
\newc{\hv}{\hat{v}}
\newc{\hk}{\hat{K}}
\newc{\ra}{\rightarrow}
\newc{\non}{\nonumber}
\newc{\ul}{\underline}
\newc{\hs}{\hspace}
\newc{\longla}{\longleftarrow}
\newc{\ts}{\textstyle}
\newc{\f}{\frac}
\newc{\df}{\dfrac}
\newc{\ovl}{\overline}
\newc{\bc}{\begin{center}}
\newc{\ec}{\end{center}}
\newc{\dg}{\dagger}
\newc{\prh}{\mbox{PR}_H}
\newc{\prq}{\mbox{PR}_q}

\begin{document}
\title{Moments of Wishart-Laguerre and Jacobi ensembles of random matrices: application to the quantum transport problem in chaotic cavities%
}
\author{Giacomo Livan
\address{Dipartimento di Fisica Nucleare e Teorica\\
Universit\`a degli Studi di Pavia\\and\\
Istituto Nazionale di Fisica Nucleare, Sezione di Pavia, \\
Via Bassi 6,
27100 Pavia, Italy 
} 
 \and
 Pierpaolo Vivo
 \address{Abdus Salam International Centre for
Theoretical Physics\\
 Strada Costiera 11,
 34151 Trieste,
 Italy}
}

\maketitle
\begin{abstract}
We collect explicit and user-friendly expressions for one-point densities of the real eigenvalues $\{\lambda_i\}$ of $N\times N$ 
Wishart-Laguerre and Jacobi random matrices with orthogonal, unitary and symplectic symmetry. Using these formulae, we compute integer moments $\tau_n=\langle\sum_{i=1}^N\lambda_i^n\rangle$ for all symmetry classes without any large $N$ approximation. In particular, our results provide exact expressions for moments of
transmission eigenvalues in chaotic cavities with time-reversal or spin-flip symmetry and supporting a finite and arbitrary number of electronic channels in the
two incoming leads. 
\end{abstract}

\tableofcontents

\section{Introduction}
We consider the Wishart-Laguerre (L) and Jacobi (J) ensembles of random matrix theory characterized by the following joint probability density (jpd) of real eigenvalues \cite{Wishart,James,muir}:
\begin{align}
\mathcal{P}_\beta^{(L)}(\lambda_1,\ldots,\lambda_N) &:=C_{N,\beta,\nu}^{(L)}\prod_{j<k}|\lambda_j-\lambda_k|^\beta \prod_{j=1}^N \lambda_j^{\frac{\beta}{2}(\nu+1)-1} e^{-\frac{1}{2}\lambda_j}\\
\mathcal{P}_\beta^{(J)}(\lambda_1,\ldots,\lambda_N) &:=C_{N,\beta,a,b}^{(J)}\prod_{j<k}|\lambda_j-\lambda_k|^\beta \prod_{j=1}^N (1-\lambda_j)^a (1+\lambda_j)^b.\label{jacobi}
\end{align}
In the above equations, $C_{N,\beta,\nu}^{(L)}$ and $C_{N,\beta,a,b}^{(J)}$ are normalization constants, while the index $\beta=1,2,4$ characterizes the symmetry class of the ensemble (orthogonal, unitary and symplectic respectively).

The Wishart-Laguerre ensemble contains covariance matrices of the form $\mathcal{W}=\mathcal{X}\mathcal{X}^\dagger$, where $\mathcal{X}$ is a $N\times M$ ($M-N:=\nu\geq 0$)
matrix with i.i.d. Gaussian entries (real, complex or quaternionic variables). The matrix $\mathcal{W}$ is symmetric and positive semidefinite, so its $N$ real eigenvalues are non-negative ($\{\lambda_j\}\geq 0$).
Originally introduced by Wishart \cite{Wishart}, matrices from this ensemble have been extensively used in
multivariate statistical data analysis~\cite{Wilks,Johnstone} with
applications in various fields ranging from meteorological
data~\cite{Preisendorfer} to finance~\cite{BP,Burda}. They are also useful when analyzing the capacity of channels with
multiple antennae and receivers~\cite{SP}, in
nuclear physics~\cite{Fyo1}, chiral quantum
chromodynamics~\cite{QCD} and also in statistical physics such as
in a class of $(1+1)$-dimensional directed polymer
problems~\cite{Johansson}. Recently, they have also
appeared in the context of knowledge networks~\cite{MZ1} and new
mathematical results
for the case $\nu<0$ have also been lately
obtained~\cite{Z2,JN}. Large deviation properties of the eigenvalues have been investigated in \cite{vivolarge,castillo,nadalmaj}, while for an excellent review
we refer to \cite{majreview}.

The Jacobi ensemble contains combinations of two $N\times N$ Wishart-Laguerre matrices $\mathcal{W}_1$ and $\mathcal{W}_2$ of the form:
\begin{equation}
\mathcal{J} = (\mathcal{W}_1-\mathcal{W}_2)(\mathcal{W}_1+\mathcal{W}_2)^{-1}
\end{equation}
and its eigenvalues are real and lie on the support $-1\leq \{\lambda_j\}\leq 1$.
Matrices distributed according to the Jacobi weight arise as $\iota)$ truncations of Haar orthogonal, unitary or symplectic matrices (for the case of 
unitary matrices, an important application arises in the theory of electronic transport 
in mesoscopic systems at low temperatures as detailed in Appendix \ref{appA}); $\iota\iota)$ as composition of projection matrices \cite{collins}.

In order to proceed, we first define a \emph{shifted} version of the Jacobi ensemble with eigenvalues between $0$ and $1$:
\begin{equation}\label{shiftedjacobi}
\mathcal{P}_\beta^{(sJ)}(\lambda_1,\ldots,\lambda_N) :=C_{N,\beta,\mathfrak{a},\mathfrak{b}}^{(sJ)}\prod_{j<k}|\lambda_j-\lambda_k|^\beta \prod_{j=1}^N \lambda_j^{\mathfrak{a}} (1-\lambda_j)^{\mathfrak{b}}.
\end{equation}
which appears more frequently in physical applications (see Appendix \ref{appA}) and numerical algorithms \cite{dumitriu,edeljac}.
By changing variables $\lambda_j=1-2 y_j$ in \eqref{jacobi}, it is easy to see that:
\begin{equation}
C_{N,\beta,a,b}^{(J)}=\frac{C_{N,\beta,a,b}^{(sJ)}}{2^{N(a+b+1)+\frac{\beta}{2}N(N-1)}}.
\end{equation}
We also define the average spectral densities $(\rho(x)=\Big\langle\frac{1}{N}\sum_{i=1}^N\delta(x-\lambda_i)\Big\rangle)$ for the three ensembles above as the marginals of their respective jpds:
\begin{align}
\rho_{N,\beta,\nu}^{(L)}(\lambda_1) &=\int_{[0,\infty]^{N-1}} d\lambda_2\cdots d\lambda_N \mathcal{P}_\beta^{(L)}(\lambda_1,\ldots,\lambda_N)\\
\rho_{N,\beta,a,b}^{(J)}(\lambda_1) &=\int_{[-1,1]^{N-1}} d\lambda_2\cdots d\lambda_N \mathcal{P}_\beta^{(J)}(\lambda_1,\ldots,\lambda_N)\\
\rho_{N,\beta,\mathfrak{a},\mathfrak{b}}^{(sJ)}(\lambda_1) &=\int_{[0,1]^{N-1}} d\lambda_2\cdots d\lambda_N \mathcal{P}_\beta^{(sJ)}(\lambda_1,\ldots,\lambda_N).
\end{align}
It follows immediately from the definition that the above densities are all normalized to $1$, and that the following relation holds between the Jacobi and the shifted-Jacobi densities:
\begin{equation}\label{reljac}
\rho_{N,\beta,a,b}^{(J)}(1-2x)=\frac{1}{2}\rho_{N,\beta,a,b}^{(sJ)}(x).
\end{equation}
The purpose of this paper is twofold:
\begin{itemize}
\item to collect and present in a user-friendly way (well-known, but somehow scattered throughout the literature) explicit formulae for the above densities for finite $N$ and all symmetry classes;
\item to use these formulae to compute integer moments of the eigenvalues $\tau_n=\langle\sum_{i=1}^N\lambda_i^n\rangle$ (where the average is taken w.r.t. any of the three jpds above).
More precisely, we define:
\begin{align}
\tau_n^{(L)}(N,\beta,\nu) &:=\int_{[0,\infty]^N}  d\lambda_1\cdots d\lambda_N \mathcal{P}_\beta^{(L)}(\lambda_1,\ldots,\lambda_N)\ \left(\sum_{i=1}^N \lambda_i^n\right)\\
\tau_n^{(J)}(N,\beta,a,b) &:=\int_{[-1,1]^N}  d\lambda_1\cdots d\lambda_N \mathcal{P}_\beta^{(J)}(\lambda_1,\ldots,\lambda_N)\ \left(\sum_{i=1}^N \lambda_i^n\right)\\
\tau_n^{(sJ)}(N,\beta,\mathfrak{a},\mathfrak{b}) &:=\int_{[0,1]^N}  d\lambda_1\cdots d\lambda_N \mathcal{P}_\beta^{(sJ)}(\lambda_1,\ldots,\lambda_N)\ \left(\sum_{i=1}^N \lambda_i^n\right).
\end{align}
One application to the case of quantum transport in chaotic cavities is detailed in Appendix \ref{appA}. Other interesting mathematical results for the Laguerre case can be found e.g. in \cite{redel}
and references therein. We are also aware that formulae for the Wishart-Laguerre and Jacobi moments for $\beta=1,2,4$ and finite $N$ have been derived by Mezzadri and Simm \cite{simm}. These are different from
the ones provided here and obtained via a different method, but equivalent.
\end{itemize}
It is easy to see that the average of any linear statistics (i.e. a quantity of the form $\mathcal{A}=\sum_{i=1}^N f(\lambda_i)$),
\begin{equation}
\langle\mathcal{A}\rangle=\int  d\lambda_1\cdots d\lambda_N \mathcal{P}_\beta(\lambda_1,\ldots,\lambda_N)\ \left(\sum_{i=1}^N f(\lambda_i)\right)
\end{equation}
 can be computed as a one-dimensional integral over the corresponding spectral 
density of the ensemble as:
\begin{equation}\label{intrho}
\Big\langle\sum_{i=1}^N f(\lambda_i)\Big\rangle=N\int dx \rho(x) f(x).
\end{equation}
The technical achievement we report in this paper is an explicit computation of this integral valid for finite matrix dimension $N$ and all three $\beta$'s for the case $f(x)=x^n$, $n\in\mathbb{N}$.
The integral \eqref{intrho} also elucidates a possible strategy to evaluate a regular asymptotic expansion of moments for large $N$, which has been recently highlighted \cite{kui} as a problem of current 
interest in the context of electronic transport in chaotic cavities (see Appendix \ref{appA}): all one has to do is to seek for a regular $(1/N)$ expansion of the macroscopic spectral density of the form:
\begin{equation}
\rho(x)=\rho^{(\infty)}(x)+\frac{1}{N}\rho^{(1)}(x)+\frac{1}{N^2}\rho^{(2)}(x)+\ldots
\end{equation}
in the spirit of high genus correlator expansions \cite{corr}, and then integrate term by term. While this program is far from completion and is thus left for future work, we show in Appendix \ref{appB} that at least the leading order of the asymptotic expansion of moments is well reproduced for the case of quantum transport in cavities with broken time-reversal symmetry ($\beta=2$).

The plan of this paper is as follows: in Section \ref{lagensemble} we consider the Wishart-Laguerre ensemble. We first summarize the densities for all three $\beta$s,
then use the integral formula \eqref{intrho} to compute integer moments for $\beta=1,2,4$ in the three subsections. Then in the last subsection we collect results from numerical
simulations for the density and moments. The same thing is done for the Jacobi ensemble in Section \ref{jacensemble}. A summary and outlook is provided in Section
\ref{conclusion}. In Appendix \ref{appA} we provide a detailed introduction to the problem of quantum transport in chaotic cavities which constitutes the main motivation for this study, while in
Appendix \ref{appB} we show that the leading order term in the expansion of the moments via spectral density is correctly reproduced.

\section{Wishart-Laguerre ensemble}\label{lagensemble}

\subsection{Spectral densities}
The spectral density for the Laguerre ensemble is known for all three symmetry classes \cite{Mehta,nagao,ghosh,kumar} and, after tedious algebraic manipulations can be cast in the form $\rho^{(L)}_{N,\beta,\nu}(x)=\frac{1}{2N}\mathcal{R}_{N,\beta,\nu}^{(L)}(x/2) $, where:
\begin{align}
\mathcal{R}_{N,1,\nu}^{(L)}(x) &=2 \mathcal{R}_{N,2,\nu}^{(L)}(2x)-\frac{\Gamma((N+1)/2)}{\Gamma((N+\nu)/2)}L_{N-1}^{(\nu)}(2x)\left\{\phi_1(x)-\phi_2(x)\right\}\label{R1lag}\\
\mathcal{R}_{N,2,\nu}^{(L)}(x) &=x^{\nu}e^{-x}\sum_{m=0}^{N-1}\frac{\Gamma(m+1)}{\Gamma(m+\nu+1)}\left(L_m^{(\nu)}(x)\right)^2 \label{R2lag}\\
\nonumber \mathcal{R}_{N,4,\nu}^{(L)}(x) &=\frac{1}{2} \mathcal{R}_{2N,2,2\nu}^{(L)}(x)-x^{2\nu}e^{-x}L_{2N}^{(2\nu)}(x)
\frac{\Gamma(N+1)}{2^{2\nu+1}\Gamma(N+\nu+1/2)}\times\\
&\times \sum_{m=0}^{N-1}
\frac{\Gamma(m+1/2)}{\Gamma(m+\nu+1)}L_{2m}^{(2\nu)}(x)\label{R4lag}
\end{align}
where:
\begin{align}
\phi_1(x) &=(2x)^\nu e^{-2x} \sum_{m=0}^{(\kappa+N-2)/2} \mathfrak{d}_m L_{2m+1-\kappa}^{(\nu)}(2x)\\
\phi_2(x) &=x^{(\nu-1)/2}e^{-x}\left[(1-\kappa)\frac{2\Gamma((\nu+1)/2,x)}{\Gamma((\nu+1)/2)}+2\kappa-1\right]\\
\mathfrak{d}_m &=\frac{\Gamma(m+1-\kappa/2)}{2^{\nu-1}\Gamma(m+(\nu-1)/2+2-\kappa/2)}.
\end{align}
In the above equations, $\kappa=N\ \mathrm{mod}\ 2$, $L_N^{(\nu)}(x)$ is a generalized Laguerre polynomial defined by the sum:
\begin{equation}\label{laguerredef}
L_n^{(\lambda)}(z)=\sum_{k=0}^n c_k(n,\lambda)z^k
\end{equation}
where:
\begin{equation}
c_k (n,\lambda)=\frac{\Gamma(\lambda+n+1)(-n)_k }{n!k!\Gamma(\lambda+k+1)}
\end{equation}
and $\Gamma(a,x)=\int_x^\infty t^{a-1}e^{-t}dt$ is the incomplete Gamma function, while $(x)_n = \Gamma(x+n)/ \Gamma(x)$ is the Pochhammer symbol.

In the following subsections, we present the results for integer moments $\tau_n^{(L)}(N,\beta,\nu)=\Big\langle\sum_{j=1}^N\lambda_j^n\Big\rangle$ 
obtained by integration (eq. \eqref{intrho}) of the densities given above. The task is most easily
accomplished by first defining the following auxiliary function:
\begin{equation}
\mathcal{Q}(r;m,\ell;\alpha):=\int_0^\infty dx\ x^r e^{-x}L_m^{(\alpha)}(x)L_\ell^{(\alpha)}(x)
\end{equation}
which is easily evaluated using \eqref{laguerredef} as:
\begin{equation}\label{Qdef}
\mathcal{Q}(r;m,\ell;\alpha)=\sum_{k=0}^m\sum_{k^\prime=0}^\ell c_k(m,\alpha)c_{k^\prime}(\ell,\alpha)\Gamma(1+r+k+k^\prime).
\end{equation}
Note that, by the orthogonality relation of Laguerre polynomials, one has:
\begin{equation}
\mathcal{Q}(\alpha;m,\ell;\alpha)=\frac{(\ell+\alpha)!}{\ell !}\delta_{m\ell}.
\end{equation}

\subsection{Moments $\beta=1$}

For $\beta=1$, the final result of the integration reads as follows:
\begin{equation}
\boxed{\tau_n^{(L)}(N,1,\nu) =2^{-n}\tau_n^{(L)}(N,2,\nu)+\mathcal{I}_2(n)+\mathcal{I}_3(n)}
\end{equation}
where:
\begin{align}
\mathcal{I}_2(n) &=-\frac{\Gamma((N+1)/2)}{2\Gamma((N+\nu)/2)}\sum_{m=0}^{(\kappa+N-2)/2}\mathfrak{d}_m\mathcal{Q}(n+\nu;N-1,2m+1-\kappa;\nu)\\
\nonumber\mathcal{I}_3(n) &=\frac{\Gamma((N+1)/2)}{2\Gamma((N+\nu)/2)} \left\{(1-\kappa)\left[\frac{2^{(3-\nu)/2}}{\Gamma((\nu+1)/2)}\mathcal{Y}_1(n)-2^{(1-\nu)/2}\mathcal{Y}_2(n)\right]+\right.\\
&\left.+\kappa\ 2^{(1-\nu)/2}\mathcal{Y}_2(n)\right\}\\
\nonumber\mathcal{Y}_1(n) &=\sum_{m=0}^{N-1} \frac{c_m (N-1,\nu) 2^{n + (\nu + 1)/2 + m}
\Gamma(n + \nu + 1 + m)}{n + (\nu + 1)/2 + m}\times\\
&\times\ _2 F_1(n + (\nu + 1)/2 + m, n + \nu + 1 + m; n + (\nu + 3)/2 + m; -1)\label{hyplag}\\
\mathcal{Y}_2(n) &=\sum_{m=0}^{N-1} c_m (N-1,\nu) 2^{n + (\nu + 1)/2 + m}
\Gamma(n + m + (\nu + 1)/2).
\end{align}
In \eqref{hyplag}, we have used the following hypergeometric function:
\begin{equation}
_2 F_1(a_1,a_2;b_1;z):=\sum_{k=0}^\infty \frac{(a_1)_k (a_2)_k}{(b_1)_k \ k!}z^k.
\end{equation}
One can check by direct inspection that $\tau_0^{(L)}(N,1,\nu)=N$ and $\tau_1^{(L)}(N,1,\nu)=N(N+\nu)$ as it should be.

\subsection{Moments $\beta=2$}

Combining \eqref{intrho} with \eqref{R2lag}, one easily obtains:
\begin{equation}
\boxed{\tau_n^{(L)}(N,2,\nu)=2^n\sum_{m=0}^{N-1}\frac{\Gamma(m+1)}{\Gamma(m+\nu+1)}\mathcal{Q}(n+\nu;m,m;\nu)}
\end{equation}
One can check by direct inspection that $\tau_0^{(L)}(N,2,\nu)=N$ and $\tau_1^{(L)}(N,2,\nu)=2N(N+\nu)$ as it should be.

\subsection{Moments $\beta=4$}
Similarly for $\beta=4$ one gets:
\begin{equation}
\boxed{\tau_n^{(L)}(N,4,\nu) =\frac{1}{2}\tau_n^{(L)}(2N,2,2\nu)-\sum_{m=0}^{N-1}\sum_{k=0}^{2N}\sum_{k^\prime=0}^{2m}
\chi_{m,k,k^\prime}}
\end{equation}
where:
\begin{align}
\chi_{m,k,k^\prime} &=f_m(N,\nu,n,k,k^\prime)c_k (2N,2\nu)c_{k^\prime} (2m,2\nu)\\
f_m(N,\nu,n,k,k^\prime) &=\frac{\Gamma(1+k+k^\prime+n+2\nu)\Gamma(N+1)\Gamma(m+1/2)}{2^{2\nu+1-n}
\Gamma(N+\nu+1/2)\Gamma(m+\nu+1)}.
\end{align}
One can check by direct inspection that $\tau_0^{(L)}(N,4,\nu)=N$ and $\tau_1^{(L)}(N,4,\nu)=4N(N+\nu)$ as it should be.

\subsection{Comparison with numerics}
\begin{figure}[ht]
\begin{center}
\includegraphics[bb=-223   180   836   612,width=\hsize]{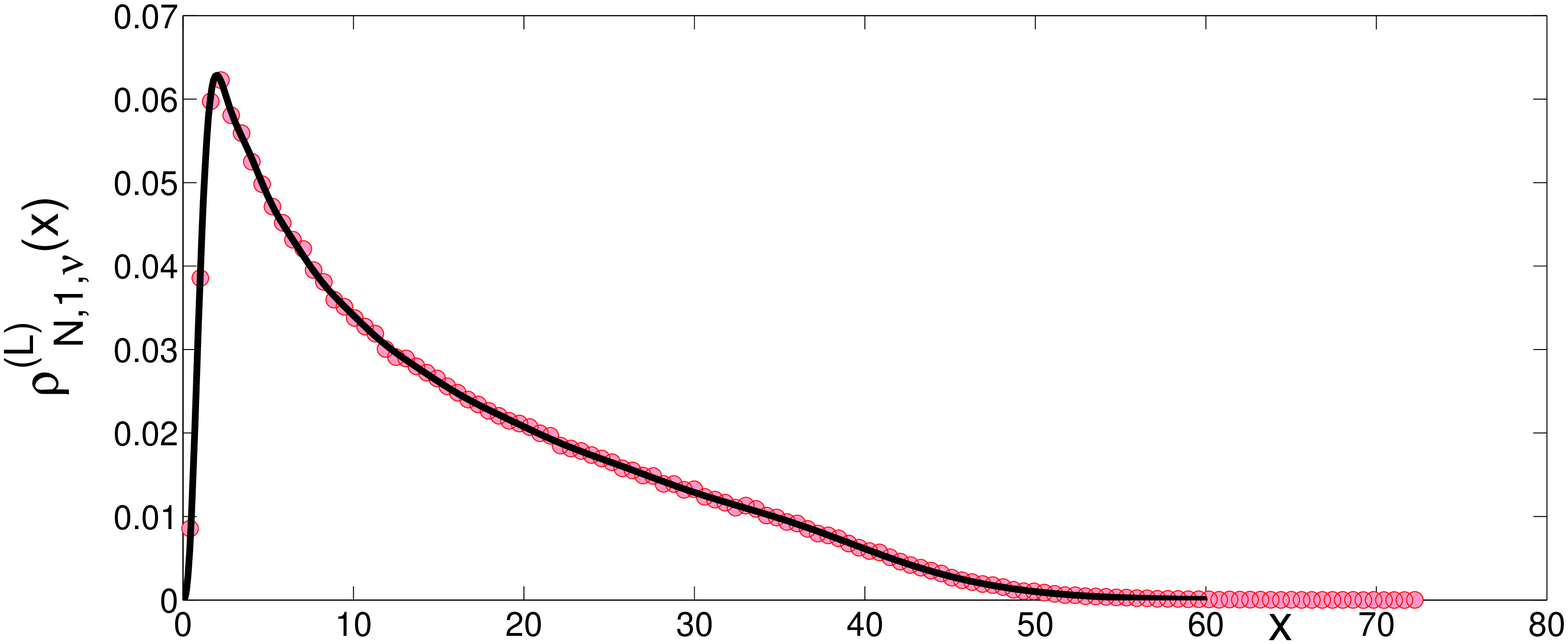}
\includegraphics[bb=-223   172   836   620,width=\hsize]{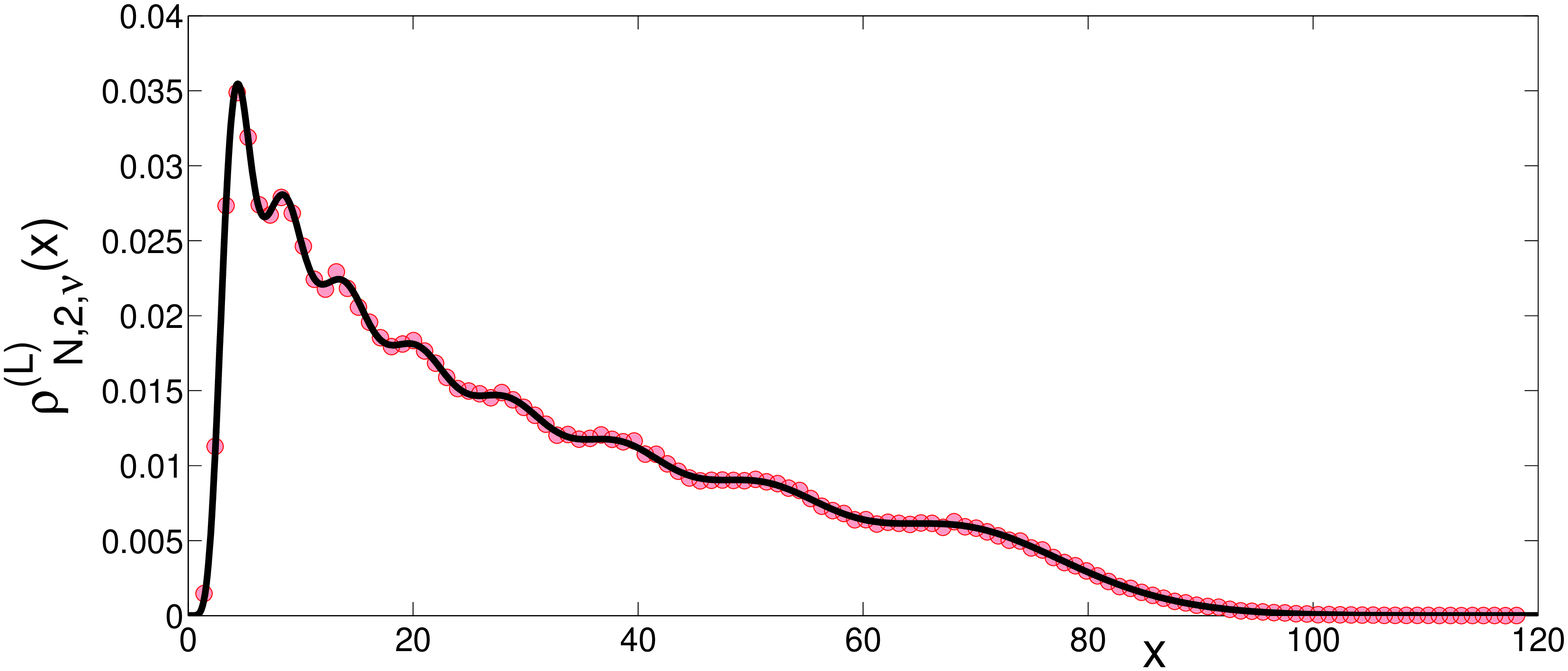}
\includegraphics[bb= -223   180   836   612,width=\hsize]{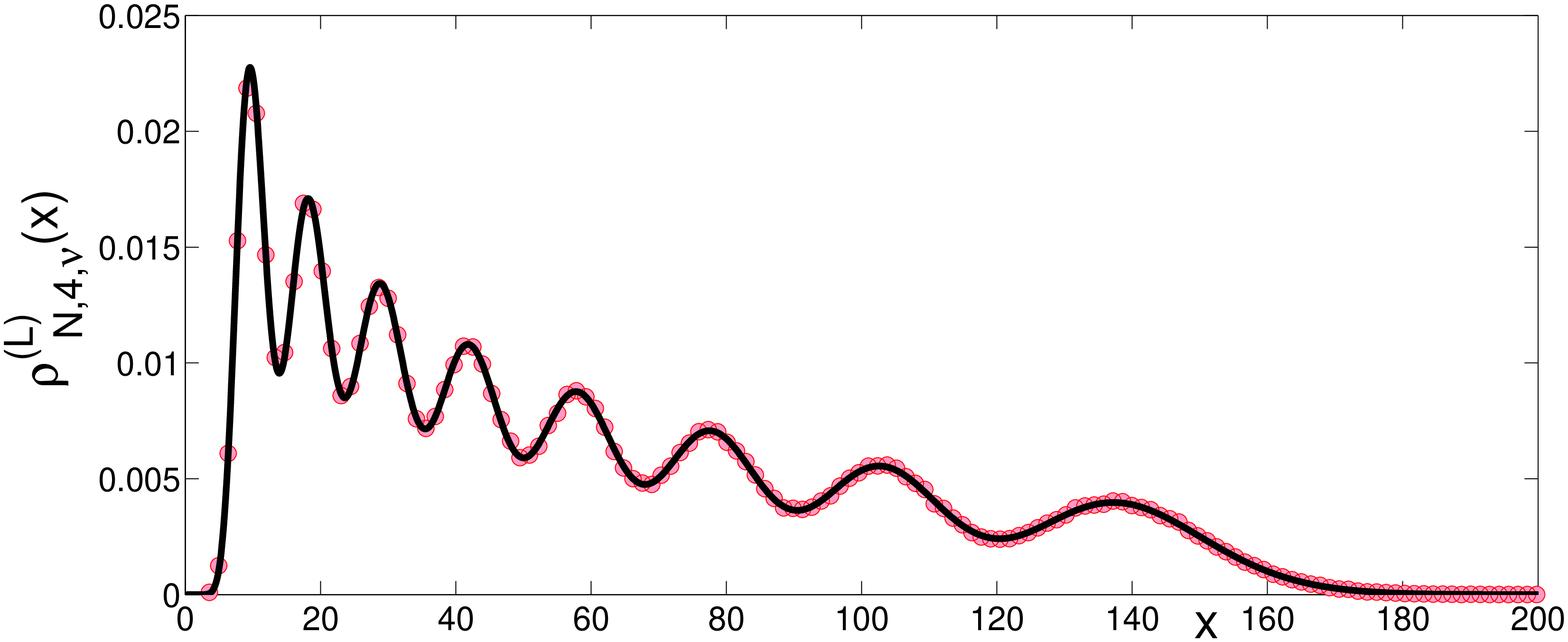}
\caption{Average spectral density of the Laguerre ensemble for $N=8,\nu=7$ and $\beta=1,2,4$ (top to bottom). Numerical diagonalization is given in red dots, while theoretical 
results are in solid black.} \label{Laguerrefig}
\end{center}
\end{figure}

In Fig. \ref{Laguerrefig} we plot the analytical formulae \eqref{R2lag}, \eqref{R1lag} and \eqref{R4lag} together with numerical diagonalization of matrices $\mathcal{W}$ from the Laguerre ensemble
with $\beta=1,2,4$ respectively\footnote{Note that for $\beta=4$ the matrices corresponding to a certain $N$ actually have size $2N\times 2N$ and thus have $2N$ real and positive eigenvalues. Only $N$
of them are distinct though, and only those must be used when carrying out numerical simulations for moments.}, obtained in \textsf{Matlab} as follows\footnote{Alternatively, one can use the tridiagonal algorithm by Dumitriu and Edelman \cite{dumitriu}.}:
\begin{align*}
\beta=1 & &\mbox{\texttt{X=randn(N,M); W = X*X';}}\\
\beta=2 & &\mbox{\texttt{X=randn(N,M)+i*randn(N,M); W = X*X';}}\\
\beta=4 && \mbox{\texttt{X = randn(N,M)+i*randn(N,M);}}\\
&& \mbox{\texttt{Y = randn(N,M)+i*randn(N,M);}}\\
  &&  \mbox{\texttt{A = [X Y; -conj(Y) conj(X)]; W = A*A';}}
\end{align*}
In Table \ref{table:nonlin1} below, we compare the analytical results for moments of the Laguerre ensemble with numerical simulations
via the \textsf{Matlab} algorithm given above.

\begin{table}[ht] 
\caption{Comparison between theory and numerics for moments of the Laguerre ensemble $(N=5,M=7\to\nu=2)$ (the numerical
results are obtained averaging over $\mathcal{O}(10^5)$ samples).} 
\centering      
\begin{tabular}{c c c c}  
\hline\hline                        
$\beta$ & $n$ & Theory & Numerics \\ [0.5ex] 
\hline
1 & 1 & 35 &  34.994\\
1 & 2 & 455 & 455.27311\\
1 & 3 & 7665 & 7660.026\\
\hline         \\           
2 & 1 & 70 & 69.981\\
2 & 2 & 1680 & 1680.238  \\   
2 & 3 & 50400 &  50409.428  \\ 
\hline \\
4 & 1 & 140  & 139.982  \\ 
4 & 2 & 6440 & 6437.074 \\ 
4 & 3 & 362880 & 362857.134 \\ [1ex]       
\hline     
\end{tabular} 
\label{table:nonlin1}  
\end{table} 

\section{Jacobi ensemble}\label{jacensemble}

\subsection{Spectral densities}
The eigenvalue densities for the Jacobi Ensemble ($x\in[-1,1]$) read \cite{ghosh,kumar}\footnote{For $\beta=1$, $N$ is even. The expression for $N$ odd is slightly more complicated \cite{ghosh}.}:

\begin{align} 
\label{density1jacobi}\rho^{(J)}_{N,1,a,b}(x) &= \frac{1}{N}\sum_{m=0}^{N/2-1} (g_{2m})^{-1} \left [ \phi_{2m}(x) \psi_{2m+1}(x) - \phi_{2m+1}(x) \psi_{2m}(x) \right ]\\
\label{density2jacobi}\rho^{(J)}_{N,2,a,b}(x) &= \frac{w_{a,b}(x)}{N} \sum_{n=0}^{N-1} (h_n^{a,b})^{-1} \left [ P_n^{(a,b)}(x) \right ]^2\\
\label{density4jacobi}\rho^{(J)}_{N,4,a,b}(x) &= \rho^{(J)}_{2N,2,2\hat{a}+1,2\hat{b}+1}(x)- P_{2N}^{(2\hat{a}+1,2\hat{b}+1)}(x) \sum_{m=0}^{N-1} K_{2N,m}^{(\hat{a},\hat{b})} P_{2m}^{(2\hat{a}+1,2\hat{b}+1)}(x)
\end{align}
where: 

\begin{align} \label{hcoeff}
h_{n}^{a,b} &= \frac{2^{a+b+1}}{2n+a+b+1} \frac{\Gamma(n+a+1) \Gamma(n+b+1)}{n! \Gamma(n+a+b+1)}\\
g_{2m} &= g_{2m+1} = h_{2m}^{2a+1,2b+1}\\
\label{phieven}\phi_{2m}(x) &= w_{a,b}(x) P_{2m}^{(2a+1,2b+1)}(x) \\
\label{psiodd}\psi_{2m+1}(x) &= w_{a+1,b+1}(x) P_{2m}^{(2a+1,2b+1)}(x)\\
\label{phiodd}\phi_{2m+1}(x) &= w_{a,b}(x) \left [ A_{2m+1} P_{2m+1}^{(2a+1,2b+1)}(x) - B_{2m-1} P_{2m-1}^{(2a+1,2b+1)}(x) \right ] \\ 
\label{psieven}\psi_{2m}(x) &= \frac{1}{2} \int_{-1}^{+1} d y \ \mathrm{sign}(x-y) \ \phi_{2m}(y) \\
\nonumber K_{N,m}^{(a,b)} &= \frac{(4m+2a+2b+3) \ \Gamma((N+2)/2) \ \Gamma((N+2a+2b+4)/2)}{2^{2a+2b+3} \ \Gamma((N+2a+2)/2) \ \Gamma((N+2b+2)/2)}\times \\
&\times\frac{\Gamma(m+1/2) \ \Gamma(m+a+b+3/2)}{\Gamma(m+a+3/2) \ \Gamma(m+b+3/2)}\\
\hat{a} &= \frac{a-2}{2}\\
\hat{b} &= \frac{b-2}{2}
\end{align}
where $\mathrm{sign}(z)=z/|z|$ and:

\begin{equation} \label{ABcoeff}
A_n = - \frac{n(n+2a+2b+2)}{2n+2a+2b+1} \ ; \ \ \ 
B_n = - \frac{(n+2a+2)(n+2b+2)}{2n+2a+2b+5} \ ; \ \ \ (B_{-1}=0).
\end{equation}
In the formulas above, $w_{a,b}(x) = (1-x)^a(1+x)^b$ is the Jacobi weight function, and $P_{n}^{(a,b)}$ is the $n-$th order Jacobi polynomial with parameters $a$ and $b$, defined as:

\begin{align} 
\nonumber P_n^{(a,b)}(x) &= \sum_{j=0}^n \binom{n+a}{j} \binom{n+b}{n-j} \left (\frac{x-1}{2} \right )^{n-j} \left ( \frac{x+1}{2} \right )^j = \\
&=\frac{1}{2^n} \sum_{j=0}^n c_j^{(n)}(a,b) (x-1)^{n-j} (x+1)^j\label{jacobiexp}
\end{align}
where we set:

\begin{equation} \label{ccoeff}
c_j^{(n)}(a,b) = \frac{\Gamma(n+a+1)}{\Gamma(j+1) \Gamma(n+a-j+1)}
\frac{\Gamma(n+b+1)}{\Gamma(n-j+1) \Gamma(b+j+1)}.
\end{equation}
We now turn to the problem of computing integer moments of Jacobi matrices. It is easy to derive first the following relations between integer moments in the shifted-Jacobi and ordinary Jacobi ensembles:
\begin{equation}\label{relationmoments1}
\boxed{\tau_n^{(sJ)}(N,\beta,\mathfrak{a},\mathfrak{b}) =2^{-n}\sum_{k=0}^n \binom{n}{k}(-1)^k \tau_k^{(J)}(N,\beta,\mathfrak{a},\mathfrak{b})}
\end{equation}
\begin{equation}\label{relationmoments2}
\boxed{\tau_n^{(J)}(N,\beta,a,b) =\sum_{k=0}^n\binom{n}{k}(-2)^k \tau_k^{(sJ)}(N,\beta,a,b). }
\end{equation}
In the following we will therefore focus on the ordinary Jacobi case.

\subsection{Moments $\beta=1$}\label{subjacobi1}

The $n$-th moment of the distribution is therefore given by:

\begin{equation} \label{nthmoment}
\boxed{\tau^{(J)}_n (N,1,a,b)  = \sum_{m=0}^{N/2-1} (g_{2m})^{-1} \left ( I_{2m,2m+1}^{(n)} - I_{2m+1,2m}^{(n)} \right )}
\end{equation}
where:

\begin{equation} \label{Idef}
I_{j,k}^{(n)} = \int_{-1}^{+1} d x \ \phi_j(x) \ \psi_k(x) \  x^n.
\end{equation}
are computed explicitly in \eqref{int1_2} and \eqref{lab} below. As it is clear from equation \eqref{nthmoment}, we only need to compute two different kinds of integrals. Let us then start from the first kind ($I_{2m,2m+1}^{(n)}$).

Using \eqref{phieven}, \eqref{psiodd} and \eqref{jacobiexp}, we therefore get (in the following the $c_j^{(n)}$ coefficients will always depend on the pair $(2a+1,2b+1)$, so we shall omit their explicit dependence on those parameters for the rest of the subsection):

\begin{align} 
\nonumber I_{2m,2m+1}^{(n)} &= \int_{-1}^{+1} d x \ \phi_{2m}(x) \ \psi_{2m+1}(x) \ x^n =\frac{1}{2^{4m}} \sum_{i,j=0}^{2m} (-1)^{4m-i-j} c_i^{(2m)} c_j^{(2m)} \times\\ 
&\times\underbrace{\int_{-1}^{+1} d x \ (1-x)^{4m+2a-i-j+1} (1+x)^{2b+i+j+1} \ x^n}_{\mathcal{L}^{(n)}(4m+2a-i-j+1,2b+i+j+1)} \label{int1_1}
\end{align}
where we have introduced the following integral:

\begin{align} 
\nonumber\mathcal{L}^{(n)}(w,z) &= \int_{-1}^{+1} d x \ (1-x)^w (1+x)^z \ x^n =2^{w+z+1} \times\\
&\times\sum_{k=0}^n \binom{n}{k} (-2)^k \mathrm{B}(w+k+1,z+1)
\end{align}\label{L_int}
introducing Euler's Beta function $\mathrm{B}(x,y)=\Gamma(x)\Gamma(y)/\Gamma(x+y)$. Inserting this result into equation \eqref{int1_1} we obtain:

\begin{align} \label{int1_2}
\nonumber \blacktriangleright I_{2m,2m+1}^{(n)} &= 2^{2(a+b+1)+1}  \sum_{i,j=0}^{2m} (-1)^{4m-i-j} c_i^{(2m)} c_j^{(2m)} \times\\
&\times\sum_{k=0}^n \binom{n}{k} (-2)^k \mathrm{B} \left (2(2m+a+1)+k-i-j,2(b+1)+i+j \right).
\end{align}
In order to compute the second type of integrals ($I_{2m+1,2m}^{(n)}$) in \eqref{nthmoment} and \eqref{Idef}, we use \eqref{phiodd}, \eqref{psieven} and \eqref{jacobiexp} to get:

\begin{align} 
\nonumber I_{2m+1,2m}^{(n)} &= \frac{1}{2^{2m+1}} \sum_{i=0}^{2m} c_i^{(2m)} \int_{-1}^{+1} d y \ w_{a,b}(y) (y-1)^{2m-i}  (y+1)^i\times\\
&\label{int2_10}\times \underbrace{\int_{-1}^{+1} d x \ \mathrm{sign}(x-y) \ \phi_{2m+1}(x) \ x^n}_{\mathcal{K}_{2m+1}^{(n)}(y)}.
\end{align}
The integral $\mathcal{K}_{2m+1}^{(n)}(y)$ can be rewritten as follows by exploiting \eqref{phiodd}:

\begin{equation} \label{Kint_1}
\mathcal{K}_{2m+1}^{(n)}(y) = A_{2m+1} \mathcal{G}_{a,b}^{(n)}(2m+1;y) - B_{2m-1} \mathcal{G}_{a,b}^{(n)}(2m-1;y)
\end{equation}
where we have:

\begin{align} 
\nonumber &\mathcal{G}_{a,b}^{(n)}(m;y)= \int_{-1}^{+1} d x \ \mathrm{sign}(x-y) \ w_{a,b}(x) \ P_m^{(2a+1,2b+1)}(x) \ x^n= \\ \nonumber
&= \frac{1}{2^m} \sum_{j=0}^m (-1)^{m-j} c_j^{(m)} \int_{-1}^{+1} d x \ \mathrm{sign}(x-y) \ (1-x)^{a+m-j} (1+x)^{b+j} \ x^n= \\ 
&= 2^{a+b+1} \sum_{j=0}^m (-1)^{m-j} c_j^{(m)} \sum_{k=0}^n \binom{n}{k} (-2)^k\ \widetilde{\mathrm{B}}_{\frac{1-y}{2}}\left ( a+m+k-j+1,b+j+1 \right)
\end{align}\label{Gint_1}
where $\widetilde{\mathrm{B}}_y(w,z) = 2 \ \mathrm{B}_y(w,z) - \mathrm{B}(w,z)$ and $\mathrm{B}_y(w,z)=\int_0^y dt\ t^{w-1}(1-t)^{z-1}$ is the incomplete Beta function.
Therefore we get:

\begin{align} 
\nonumber &\mathcal{K}_{2m+1}^{(n)}(y) = 2^{a+b+1} \sum_{k=0}^n \binom{n}{k} (-2)^k\times \\ \nonumber
&\times \left [ A_{2m+1} \sum_{j=0}^{2m+1} (-1)^{2m-j+1} c_j^{(2m+1)} \ \widetilde{\mathrm{B}}_{\frac{1-y}{2}} \Big ( 2(m+1)+a+k-j,b+j+1 \Big ) \right. \\ 
&- \left. B_{2m-1} \sum_{j=0}^{2m-1} (-1)^{2m-j-1} c_j^{(2m-1)} \ \widetilde{\mathrm{B}}_{\frac{1-y}{2}} \Big ( 2m+a+k-j,b+j+1 \Big ) \right ].
\end{align}\label{Kint_2}
Eventually, inserting this result into \eqref{int2_10} and computing the integral in $d y$, we obtain

\begin{align} 
\nonumber & \blacktriangleright I_{2m+1,2m}^{(n)} =\frac{2^{a+b}}{2^{2m}} \sum_{i=0}^{2m} c_i^{(2m)} \sum_{k=0}^n \binom{n}{k} 
(-2)^k\times \\ \nonumber 
&\times \left [ A_{2m+1} \sum_{j=0}^{2m+1} (-1)^{2m-j+1} c_j^{(2m+1)} \ \Omega_{a,b} \Big ( 2m-i,i;2(m+1)+a+k-j,b+j+1 \Big ) \right. \\ 
&\label{lab} - \left. B_{2m-1} \sum_{j=0}^{2m-1} (-1)^{2m-j-1} c_j^{(2m-1)} \ \Omega_{a,b} \Big (2m-i,i;2m+a+k-j,b+j+1 \Big ) \right ]
\end{align}
where we have:

\begin{align} 
\nonumber &\Omega_{a,b}(h,\ell;w,z) = \int_{-1}^{+1} d x \ w_{a,b}(x) (x-1)^h (x+1)^\ell \ \widetilde{\mathrm{B}}_{\frac{1-x}{2}}(w,z) =  \\ \nonumber  
&=(-1)^h \ 2^{a+b+h+\ell+1} \left [ w^{-1} \ \mathrm{B} \Big (1+b+\ell,1+a+h+w \Big )\times\right.\\
&\nonumber\times ~_3 F_2 \Big (w,1+a+h+w,1-z;1+w,2+a+b+h+\ell+w;1 \Big)  \\ 
&-\left. \mathrm{B}(w,z) \ \mathrm{B} \Big (1+a+h,1+b+\ell \Big ) \right ].\label{Omegasquare}
\end{align}
In \eqref{Omegasquare}, we have used the following generalized hypergeometric function:
\begin{equation}
_3 F_2(a_1,a_2,a_3;b_1,b_2;z):=\sum_{k=0}^\infty \frac{(a_1)_k (a_2)_k (a_3)_k}{(b_1)_k (b_2)_k\ k!}z^k.
\end{equation}

\subsection{Moments $\beta=2$}

We wish now to compute the generic $n-$th moment of the density in \eqref{density2jacobi}, which reads

\begin{equation} \label{moment}
\tau_n^{(J)}(N,2,a,b) = \left \langle \sum_{i=1}^N x_i^n \right \rangle = N\int_{-1}^{+1} d x \ \rho^{(J)}_{N,2,a,b}(x) \ x^n.
\end{equation}
By making use of the representation of Jacobi polynomials in \eqref{jacobiexp}, the $n-$th moment in \eqref{moment} can be written as follows (throughout all the present subsection the $c_j^{(k)}$ coefficients will depend on the pair $(a,b)$):

\begin{align} \label{moment2}
\nonumber\tau_n^{(J)}(N,2,a,b) &= \sum_{k=0}^{N-1} (h_k^{a,b})^{-1} \int_{-1}^{+1}  dx \ w_{a,b}(x) \left [ P_k^{(a,b)}(x) \right ]^2 x^n =\\
&= \sum_{k=0}^{N-1} \sum_{i,j=0}^k \frac{c_i^{(k)} c_j^{(k)}}{2^{2k} h_k^{a,b}} I^{(n)}(k,i,j,a,b)
\end{align}
where:

\begin{equation} \label{Iint2}
I^{(n)}(k,i,j,a,b) = (-1)^{2k-i-j} \int_{-1}^{+1}  dx \ (1-x)^{2k+a-i-j} (1+x)^{b+i+j} x^n
\end{equation}
and can be computed by changing variables and setting $y = (1-x)/2$. When doing so, one obtains:

\begin{equation} \label{Iint3}
I^{(n)}(k,i,j,a,b) = (-1)^{2k-i-j} 2^{2k+a+b+1} \int_0^1 dy \ y^{2k+a-i-j} (1-y)^{b+i+j} (1-2y)^n,
\end{equation}
and expanding $(1-2y)^n$ according to the binomial Theorem, one gets:

\begin{align} \label{Iint4}
\nonumber I^{(n)}(k,i,j,a,b) &= (-1)^{2k-i-j} 2^{2k+a+b+1} \sum_{\ell=0}^n (-2)^\ell \binom{n}{\ell}\times\\
&\times \int_0^1 dy \ y^{2k+a+\ell-i-j} (1-y)^{b+i+j}.
\end{align}
The integral appearing in the previous equation is of the following kind

\begin{equation} \label{gammaint}
\int_0^1 dt \ t^{x-1} (1-t)^{y-1} = \mathrm{B}(x,y).
\end{equation}
Thus we get:

\begin{align}
\nonumber I^{(n)}(k,i,j,a,b) &= (-1)^{2k-i-j} 2^{2k+a+b+1} \sum_{\ell=0}^n (-2)^\ell \binom{n}{\ell} \times\\
&\times\mathrm{B}(2k+a+\ell-i-j+1,b+i+j+1).
\end{align}
Plugging this result in equation \eqref{moment2}, one can easily see that the $n-$th moment has the compact expression:

\begin{equation} \label{moment3}
\boxed{\tau_n^{(J)}(N,2,a,b) =\sum_{k=0}^{N-1} \sum_{i,j=0}^k \sum_{\ell=0}^n \mathfrak{y}_{k,i,j,\ell} \mathrm{B}(2k+a+\ell-i-j+1,b+i+j+1)}
\end{equation}
where:
\begin{equation}
\mathfrak{y}_{k,i,j,\ell}=\frac{(-1)^{2k-i-j+\ell} c_i^{(k)} c_j^{(k)} 2^{a+b+1+\ell} }{h_k^{a,b}}\binom{n}{\ell}.
\end{equation}
After inserting \eqref{moment3} into \eqref{relationmoments1}, we have checked that the special case $\mathfrak{a}=\mu,\mathfrak{b}=0$ for $\tau_n^{(sJ)}(N,2,\mathfrak{a},\mathfrak{b})$ indeed agrees with 
eq. (16) in \cite{vivovivo} and with eq. (13) in \cite{novaes2} as it should (see Appendix \ref{appA} for details).

\subsection{Moments $\beta=4$}\label{subjacobi4}
The $n$-th moment of the density in \eqref{density4jacobi} reads:

\begin{equation} \label{sympmoment}
\boxed{\tau^{(J)}_n(N,4,a,b) = \frac{1}{2} \tau^{(J)}_n(2N,2,2\hat{a}+1,2\hat{b}+1) - \frac{1}{2} \sum_{m=0}^{N-1} \Theta_{N,m}}
\end{equation}
where:

\begin{align} \label{Yint1}
\Theta_{N,m} &= K_{2N,m}^{(2\hat{a}+1,2\hat{b}+1)} Y_{2N,2m}^{(n)}(2\hat{a}+1,2\hat{b}+1)\\
Y_{k,\ell}^{(n)}(a,b) &= \int_{-1}^{+1} dx \ w_{a,b}(x) \ P_k^{(a,b)}(x) \ P_\ell^{(a,b)}(x) \ x^n.
\end{align}
Exploiting the Jacobi polynomial expansion \eqref{jacobiexp} and the binomial Theorem we get (again, we omit the explicit dependence of the $c_i^{(k)}$ coefficients on the pair $(a,b)$):

\begin{align} \label{Yint2}
\nonumber Y_{k,\ell}^{(n)}(a,b) &= \frac{1}{2^{k+\ell}} \sum_{i=0}^k \sum_{j=0}^\ell (-1)^{k+\ell-i-j} c_i^{(k)} \ c_j^{(\ell)} \times\\
\nonumber &\times \int_{-1}^{+1}  dx \ (1-x)^{a+k+\ell-i-j} \ (1+x)^{b+i+j} \ x^n= \\ 
\nonumber &= 2^{a+b+1} \sum_{i=0}^k \sum_{j=0}^\ell \sum_{s=0}^n \binom{n}{s} (-1)^{k+\ell-i-j} (-2)^s c_i^{(k)} \ c_j^{(\ell)} \times\\
&\times \mathrm{B}(a+k+\ell+s-i-j+1,b+i+j+1).
\end{align}

\subsection{Comparison with numerics}
\begin{figure}[ht]
\begin{center}
\includegraphics[bb=-252   149   865   641,width=0.95\hsize]{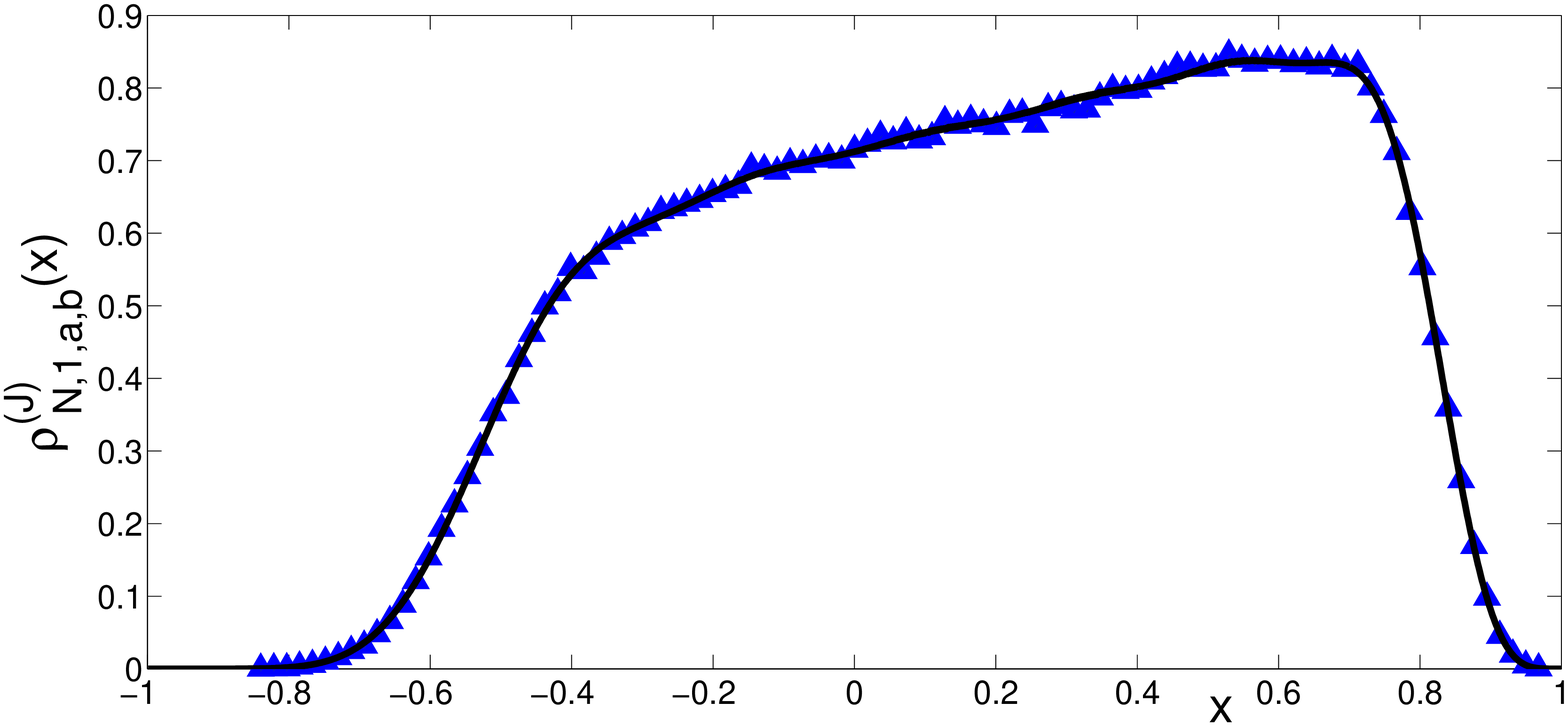}
\includegraphics[bb=-252   150   865   642,width=0.95\hsize]{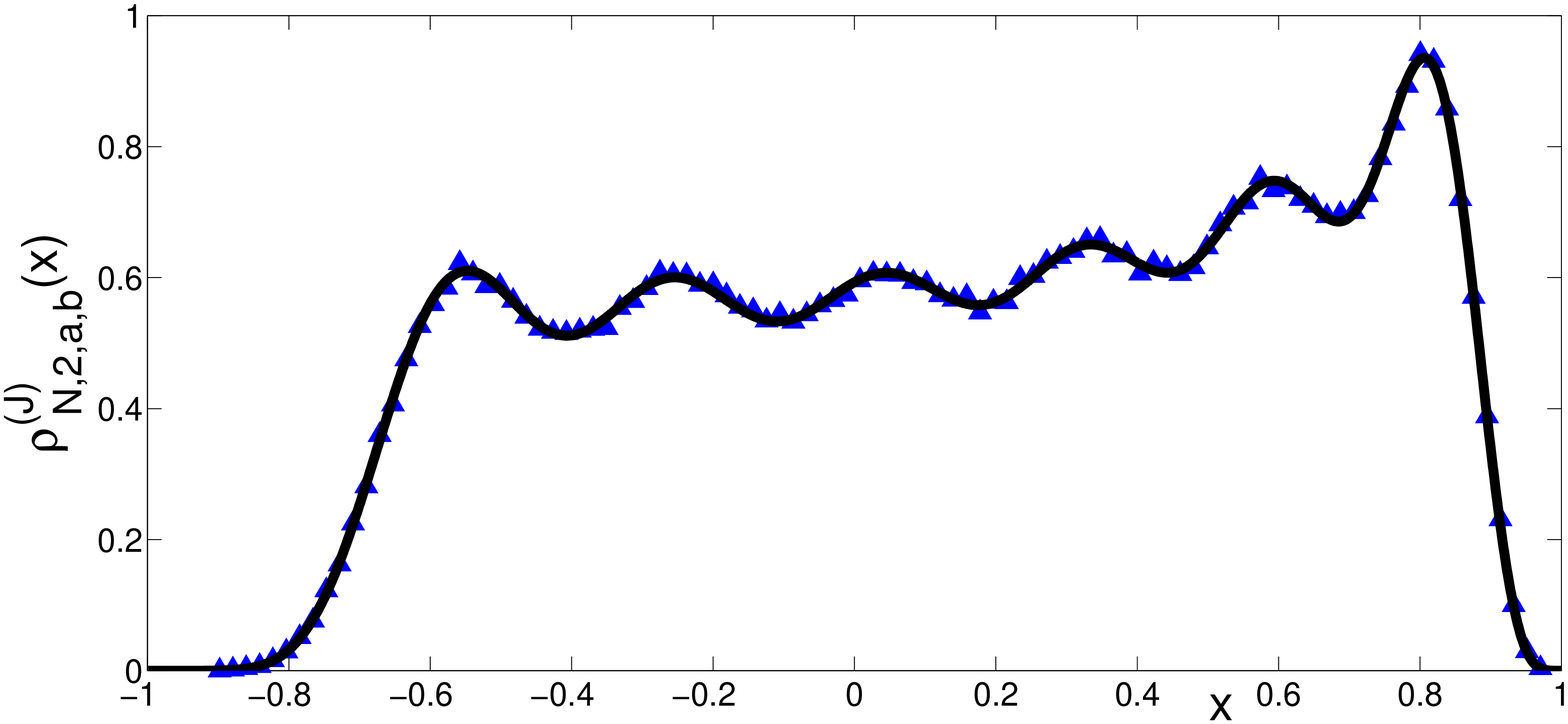}
\includegraphics[bb= -252   158   865   634,width=0.95\hsize]{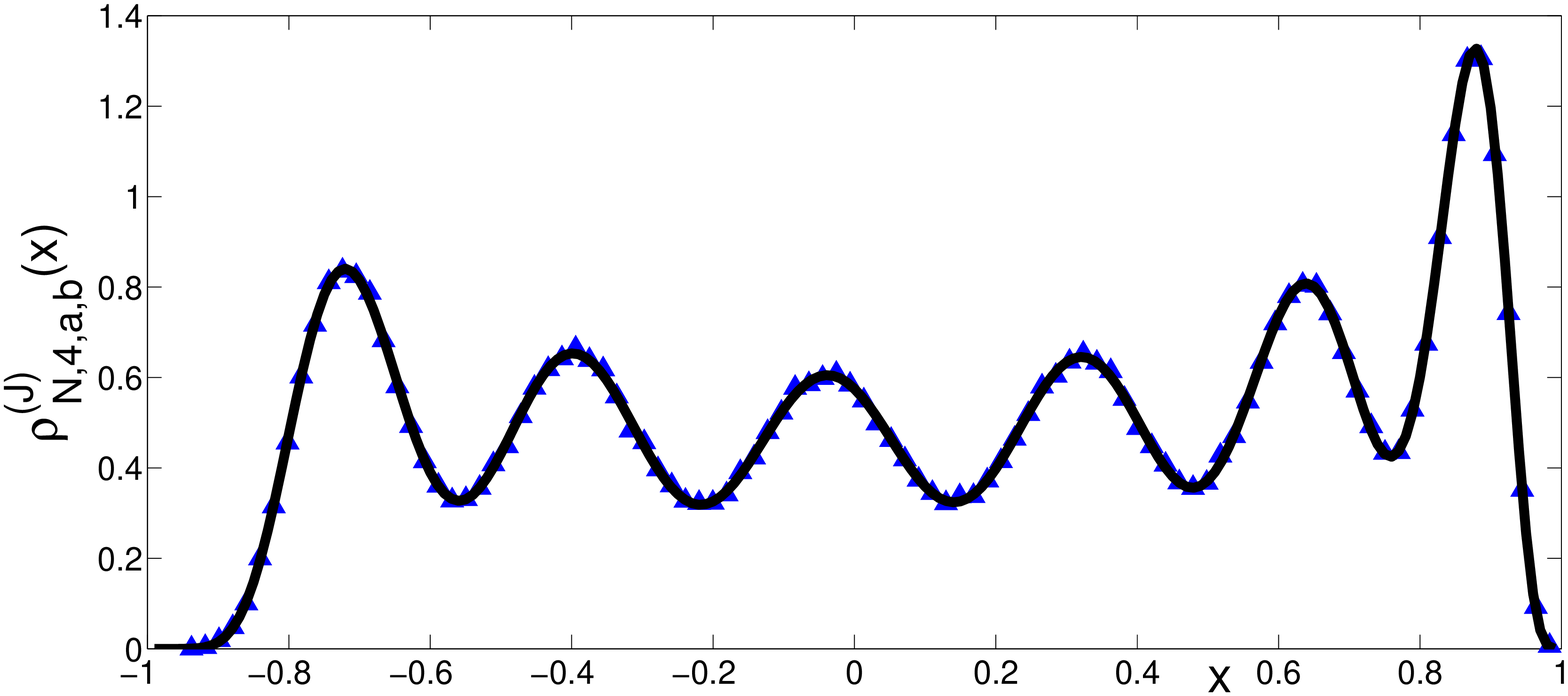}
\caption{Average spectral density of the Jacobi ensemble for $N=6,a=5,b=9$ and $\beta=1,2,4$ (top to bottom). Numerical diagonalization is given in blue triangles, while theoretical 
results are in solid black.} \label{Jacobifig}
\end{center}
\end{figure}

In Fig. \ref{Jacobifig} we plot the analytical formulae \eqref{density1jacobi}, \eqref{density2jacobi} and \eqref{density4jacobi} together with numerical diagonalization of matrices $\mathcal{J}$ from the Jacobi ensemble
with $\beta=1,2,4$ respectively, obtained in \textsf{Matlab} using the algorithm by Edelman and Sutton \cite{edeljac}.

In the following tables, we compare the analytical results for moments of the Jacobi and shifted-Jacobi ensembles with numerical simulations
via the algorithm by Edelman and Sutton \cite{edeljac}.

\begin{table}[ht] 
\caption{Comparison between theory and numerics for moments of the Jacobi ensemble $(N=6,a=5,b=9)$ (the numerical
results are obtained averaging over $\mathcal{O}(10^5)$ samples).} 
\centering      
\begin{tabular}{c c c c}  
\hline\hline                        
$\beta$ & $n$ & Theory & Numerics \\ [0.5ex] 
\hline
1 & 1 & 1.1428 &  1.1454\\
1 & 2 & 1.1536 & 1.1573\\
1 & 3 & 0.5108 & 0.5125\\
\hline         \\           
2 & 1 & 0.9230 & 0.9259\\
2 & 2 & 1.4872 & 1.4892  \\   
2 & 3 & 0.5470 & 0.5485  \\ 
\hline         \\
4 & 1 & 0.6666 & 0.6674 \\
4 & 2 & 1.9300 & 1.9294 \\
4 & 3 & 0.5602 & 0.5602 \\ [1ex]       
\hline     
\end{tabular} 
\label{table:nonlin}  
\end{table}

\begin{table}[ht] 
\caption{Comparison between theory and numerics for moments of the shifted-Jacobi ensemble $(N=6,\mathfrak{a}=5,\mathfrak{b}=9)$ (the numerical
results are obtained averaging over $\mathcal{O}(10^5)$ samples).} 
\centering      
\begin{tabular}{c c c c}  
\hline\hline                        
$\beta$ & $n$ & Theory & Numerics \\ [0.5ex] 
\hline
1 & 1 & 2.4286 &  2.4300\\
1 & 2 & 1.2170 & 1.2169\\
1 & 3 & 0.6902 & 0.6902\\
\hline         \\           
2 & 1 & 2.5385 & 2.5390\\
2 & 2 & 1.4103 & 1.4102  \\   
2 & 3 & 0.8932 & 0.8919  \\ 
\hline 	       \\
4 & 1 & 2.6667 & 2.6669 \\
4 & 2 & 1.6492 & 1.6493 \\
4 & 3 & 1.1605 & 1.1612 \\ [1ex]       
\hline     
\end{tabular} 
\label{table:nonlin}  
\end{table}

\section{Conclusions}\label{conclusion}
In conclusions, we have first collected well-known formulae for the spectral density of Laguerre and Jacobi ensembles of random matrices with orthogonal, unitary and symplectic symmetry.
We feel that this paper might serve as a quick reference point for formulae that might be hard to dig out in the literature. Using these results, we have computed the average of integer
moments, reducing the complexity from a $N$-fold integration to a single integral over the spectral density. In all cases, expressions different from ours and derived through different methods already exist \cite{simm,vivovivo,novaes2} (see also Appendix \ref{appA}). It would be interesting to prove mathematically the equivalence of various formulae which are now available for the same objects.
The present paper offers a different and possibly simpler way of deriving transport moments
for all symmetry classes and finite number of open electronic channels. The obtained results have been checked numerically with high accuracy, and the corresponding \textsf{Matlab} codes are freely available 
on demand.
\newline\\
{\bf Acknowledgments:} we gratefully acknowledge useful correspondence with Saugata Ghosh, Marcel Novaes, Francesco Mezzadri, Gernot Akemann and Dima Savin. We are indebted to Akhilesh Pandey and Santosh Kumar for valuable correspondence
and for sending us their preprint \cite{kumar} before publication.

\appendix
\section{Electronic transport in open cavities}\label{appA}

A cavity of submicron dimensions etched in a semiconductor can be connected to the external world by two leads supporting $N_1$
and $N_2$ electronic channels. A voltage difference $V$ applied between the two leads lets an electronic current flow through the cavity,
whose intensity presents time-dependent
fluctuations which persist down to zero temperature
\cite{beenakker}. These are associated with the granularity of
the electron charge $e$. Typical phenomena
observed in experiments include weak localization
\cite{chang}, universality in conductance fluctuations
\cite{marcus} and constant Fano factor \cite{oberholzer}.
According to the Landauer-B\"{u}ttiker scattering approach
\cite{beenakker,landauer,buttikerPRL}, the statistics of quantum transport observables depend in a rather simple way on the scattering matrix $\mathcal{S} $ of the cavity.
This is a unitary $N_0\times N_0$ matrix (where $N_0=N_1+N_2$) which relates the wave function
coefficients of the incoming and outgoing electrons in a natural basis:
\begin{equation}\label{ScatteringMatrix S}
  \mathcal{S}=
  \begin{pmatrix}
    \mathbf{r} & \mathbf{t}^\prime \\
    \mathbf{t} & \mathbf{r}^\prime
  \end{pmatrix}.
\end{equation}
The transmission ($\mathbf{t},\mathbf{t}^\prime$) and reflection
$(\mathbf{r},\mathbf{r}^\prime)$ blocks are submatrices encoding the
transmission and reflection coefficients among different channels\footnote{($\mathbf{t},\mathbf{t}^\prime$) are respectively
of size $N_1\times N_2$ and $N_2\times N_1$, while $(\mathbf{r},\mathbf{r}^\prime)$ are of size $N_2\times N_2$ and $N_1\times N_1$.}.
The eigenvalues of the hermitian transport matrix $\mathbf{T}=\mathbf{t} \mathbf{t}^\dagger$ are of primary importance: for
example, the dimensionless conductance and the shot noise are
given respectively by $G=\Tr(\mathbf{T})$ \cite{landauer} and
$P=\Tr[\mathbf{T}(\mathbf{1}-\mathbf{T})]$ \cite{ya}, while other transport properties are encoded in higher moments $\tau_n=\Tr[\mathbf{T}^n]$.

When the classical electronic motion inside the cavity can be regarded as chaotic, Random Matrix Theory has been very successful in describing the statistics of universal fluctuations in such systems:
the scattering matrix $\mathcal{S}$ is assumed to be drawn at random from a suitable
ensemble of matrices, with the overall constraint of
unitarity \cite{muttalib,stone,mellopereira}. A
maximum entropy approach (under the assumption of ballistic
point contacts \cite{beenakker}) forces the probability distribution of
$\mathcal{S}$ to be uniform within the unitary group, i.e. $\mathcal{S}$ belongs to
one of Dyson's Circular Ensembles \cite{Mehta,Dys:new}.

The uniformity of $\mathcal{S}$ within the unitary group induces the 
following remarkably simple jpd of transmission eigenvalues $\{T_i\}$ of the
matrix $\mathbf{T}$ \cite{beenakker,mellopereira,forrcond}:
\begin{equation}\label{jpdcond}
 \mathcal{P}_\beta^{(\mathbf{T})}(T_1,\ldots,T_N)\propto\prod_{j<k} |T_j-T_k|^\beta \prod_{i=1}^N T_i^{\frac{\beta}{2}(\mu+1)-1}
\end{equation}
with $N=\min(N_1,N_2)$, $\mu = |N_1-N_2|$ and $\beta=1,2,4$ in case of preserved time-reversal symmetry, broken time-reversal symmetry and spin-flip symmetry respectively.
The jpd \eqref{jpdcond} is precisely of the shifted-Jacobi form \eqref{shiftedjacobi} with $\mathfrak{a}=\frac{\beta}{2}(\mu+1)-1$ and $\mathfrak{b}=0$.

The transmission eigenvalues $T_i$ are thus correlated real random variables
between $0$ and $1$, whose jpd \eqref{jpdcond} in principle allows for a complete characterization of statistical properties of experimental observables.
For most recent analytical results, we refer to \cite{kumar,vivovivo,novaes2,novaes1,brouwer,savin,savin2,luque,sommers,savinnew,Kanz,vivoPRL}.

In particular, the study of higher moments of the transmission matrix $\tau_n^\star=\Tr[\mathbf{T}^n]$ has recently seen many analytical progresses\footnote{In our notation,
$\tau_n^\star \equiv \tau_n^{(sJ)}(N,2,\mu,0)$.} \cite{simm,kui,kumar,vivovivo,novaes2,novaes1,savin2,luque,vivoPRL}. In particular, we now have two different (but equivalent) formulae for higher moments for $\beta=2$ and arbitrary $N_1,N_2$:
\begin{align}
\tau_n^\star &=\sum_{p=0}^{N-1} (2p+\mu+1)\sum_{k,\ell=0}^p \frac{g_p(k)g_p(\ell)}{\mu+n+k+\ell+1}, &\quad\mbox{see }\cite{vivovivo}\\
\tau_n^\star &=\sum_{p=0}^{n-1}\frac{(-1)^p}{n!}\binom{n-1}{p}\frac{(N_1-p)_n (N_2-p)_n}{(N_1+N_2-p)_n}, &\quad\mbox{see }\cite{novaes2} \label{nov1}
\end{align}
where:
\begin{equation}
g_p(\kappa):=(-1)^\kappa\binom{p}{\kappa}\binom{p+\mu+\kappa}{\mu+\kappa}.
\end{equation}
Novaes \cite{novaes1} has also computed the leading $\mathcal{O}(N)$ term in
the {\em large} $N_1,N_2\gg 1$ expansion of the moments $\tau^{(sJ)}_n(\infty,2,\mu,0)$ (Eq. \eqref{nov1}) for the quantum transport problem at $\beta=2$. His formula in our notation reads:
\begin{equation}\label{novaes}
\tau^{(sJ)}_n(\infty,2,\mu,0)\sim(N_1+N_2)\sum_{p=1}^n \binom{n-1}{p-1}(-1)^{p-1}c_{p-1}\xi^p
\end{equation}
where $c_p=\frac{1}{p+1}\binom{2p}{p}$ and $\xi=N_1 N_2/(N_1+N_2)^2$. In the next Appendix, we will show how this result (for the case $\mu\sim\mathcal{O}(1)\to\xi=1/4$) can be derived by integration of the asymptotic spectral density of shifted Jacobi ensemble with $\mathfrak{a}=\mathfrak{b}\sim\mathcal{O}(1)$ when $N\to\infty$. 

\begin{figure}[ht]
\begin{center}
\includegraphics[bb= -249   150   862   642,width=\hsize]{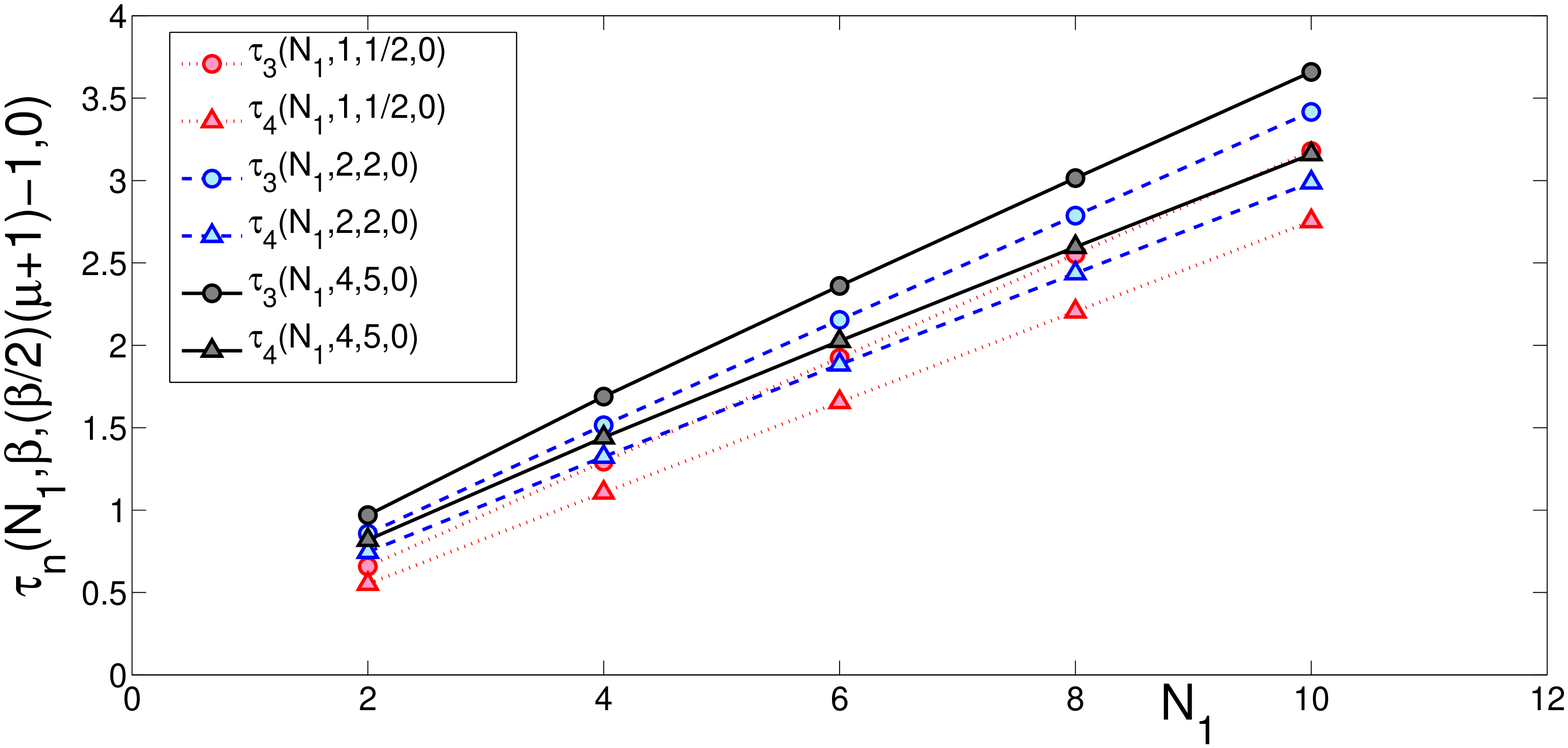}
\caption{Transport moments $\tau_n(N_1,\beta,(\beta/2)(\mu+1)-1,0)$ for $n=3$ (circles) and $n=4$ (triangles) as a function of $N_1$. The number of channels $N_2=N_1+2$ is held fixed.
The results are for $\beta=1$ (short-dashed red), $\beta=2$ (long-dashed blue) and $\beta=4$ (solid black).} \label{Jacobi_quantum}
\end{center}
\end{figure}

In Fig. \ref{Jacobi_quantum} we plot the transport moments $\tau_n(N_1,\beta,(\beta/2)(\mu+1)-1,0)$ for $n=3$ and $n=4$ as a function of $N_1$, while the number of channels $N_2=N_1+2$ is held fixed. All cases yield expressions formally different from others \cite{simm} (see equation \eqref{relationmoments1} and the equations in subsections \ref{subjacobi1} and \ref{subjacobi4}), but equivalent.

\section{Large $N$ asymptotics for moments}\label{appB}
The general integral formula \eqref{intrho} offers a neat and explicit way to obtain the large $N$ asymptotic behavior for moments and possibly 
the full $1/N$ expansion as follows. Suppose that the spectral density of the ensemble under discussion admits the following series expansion:
\begin{equation}
\rho(x)=\rho^{(\infty)}(x)+\frac{1}{N}\rho^{(1)}(x)+\frac{1}{N^2}\rho^{(2)}(x)+\ldots
\end{equation}
then the corresponding series for moments can be formally obtained as:
\begin{equation}\label{developmenttau}
\tau_n=N\int dx \rho^{(\infty)}(x) x^n+\int dx\rho^{(1)}(x) x^n+\frac{1}{N}\int dx\rho^{(2)}(x)x^n+\ldots
\end{equation}
A first check of the validity of the putative formula \eqref{developmenttau} goes as follows. Novaes \cite{novaes1} has computed the leading $\mathcal{O}(N)$ term in
the {\em large} $N_1,N_2\gg 1$ expansion of the moments $\tau^{(sJ)}_n(\infty,2,\mu,0)$ for the quantum transport problem in chaotic cavities with unitary simmetry supporting $N_1,N_2$ electronic channels in the leads ($\mu=|N_1-N_2|$ and $N=\min(N_1,N_2)$), see eq. \eqref{novaes}. 
Now, the $N\to\infty$ limit for the spectral density of the ordinary Jacobi ensemble with $a,b\sim\mathcal{O}(1)$ reads \cite{ghosh}:
\begin{equation}\label{rhoinfty}
\rho^{(J)}_{\infty,\beta,a,b}(x)=\frac{1}{\pi\sqrt{1-x^2}}
\end{equation}
i.e. it is independent of $a,b,\beta$.
Using \eqref{reljac} we get:
\begin{equation}
\rho_{\infty,\beta,\mathfrak{a},\mathfrak{b}}^{(sJ)}(x)  =\frac{2}{\pi\sqrt{1-(1-2x)^2}}
\end{equation}
and using \eqref{developmenttau}, the $\mathcal{O}(N)$ term in the expansion of the $n$-th moment of the $\beta=2$ shifted-Jacobi ensemble should read:
\begin{align}
\nonumber\lim_{N\to\infty}\frac{1}{N}\tau^{(sJ)}_n(N,2,\mathfrak{a},\mathfrak{b}) &=\int_{0}^1 dx\frac{2x^n}{\pi\sqrt{1-(1-2x)^2}}\\
&=\frac{\Gamma(n+1/2)}{\sqrt{\pi}\Gamma(1+n)}.\label{tauzero}
\end{align}
Eq. \eqref{tauzero} should then be compared with \eqref{novaes} for a special value of $\xi$, namely the one that guarantees that parameters $\mathfrak{a},\mathfrak{b}$ are finite (i.e. of $\mathcal{O}(1)$) when $N\to\infty$
(otherwise eq. \eqref{rhoinfty} would not hold). Since $\mathfrak{a}\equiv\mu=N_1-N_2$, we need to impose that this difference is finite when $N_1,N_2\to\infty$. This immediately leads to $\xi=1/4$
and $N_1+N_2=2N$. Inserting these values into \eqref{novaes}, we indeed find the identity:
\begin{equation}
2\sum_{p=1}^n \binom{n-1}{p-1}(-1)^{p-1}c_{p-1} 4^{-p}=\frac{\Gamma(n+1/2)}{\sqrt{\pi}\Gamma(1+n)}.
\end{equation}
Therefore, we have proven that a simple one-dimensional integral over the asymptotic spectral density of the shifted-Jacobi ensemble correctly reproduces Novaes' formula
for the asymptotics of transport moment (obtained by a totally different method) when the difference between the number of electronic channels in the two leads is of $\mathcal{O}(1)$.


\begin{thebibliography}{99}



\bibitem{Wishart} J. Wishart, The generalised product moment distribution in samples from a normal multivariate population, {\em Biometrika} {\bf 20A}, 32 (1928).

\bibitem{James} A. T. James, Distribution of matrix variates and latent roots derived from normal
samples, {\em Ann. Math. Stat.} {\bf 35}, 475 (1964).

\bibitem{muir} R. J. Muirhead, {\em Aspects of Multivariate Statistical Theory}, (2005 John Wiley \& Sons, Inc.).

\bibitem{Wilks} S. S. Wilks, {\em Mathematical Statistics} (1962 John Wiley \& Sons, New York).

\bibitem{Johnstone} I. M. Johnstone, On the distribution of the largest eigenvalue in principal components analysis, {\em Ann. Statist.} {\bf 29}, 295 (2001).

\bibitem{Preisendorfer} R. W. Preisendorfer, {\em Principal Component Analysis in Meteorology
and Oceanography} (1988 Elsevier, New York).

\bibitem{BP} J.-P. Bouchaud and M. Potters, {\em Theory of Financial Risks} (2001 Cambridge University
Press, Cambridge); L. Laloux, P. Cizeau, J.-P. Bouchaud, and M. Potters, Noise dressing of financial correlation matrices, {\em Phys. Rev. Lett.} {\bf 83}, 1467 (1999). 

\bibitem{Burda} Z. Burda and J. Jurkiewicz, 
Signal and noise in financial correlation matrices, {\em Physica A} {\bf 344}, 67 (2004);
Z. Burda, J. Jurkiewicz, and B. Wac\l aw, Eigenvalue density of empirical covariance matrix for correlated samples, {\em Acta Physica Polonica B} {\bf
36}, 2641 (2005).

\bibitem{SP} A. M. Sengupta and P. P. Mitra, Distributions of singular values for some random matrices, {\em Phys. Rev. E} {\bf 60}, 3389 (1999); E. Telatar,
Capacity of multi-antenna Gaussian channels, {\em European Transactions on Telecommunications} {\bf 10}, 585 (1999); P. Kazakopoulos, P. Mertikopoulos, A. L. Moustakas, and G. Caire, 
Living at the edge: a large deviations approach to the outage MIMO capacity, to appear in {\em IEEE Transactions on Information Theory}, [arXiv:0907.5024] (2009).

\bibitem{Fyo1} Y. V. Fyodorov and H.-J. Sommers, Statistics of resonance poles, phase shifts and time delays in quantum chaotic scattering: Random matrix approach for systems with broken time-reversal invariance, {\em J. Math. Phys.} {\bf 38}, 1918 (1997); Y. V. Fyodorov
and B. A. Khoruzhenko, Systematic analytical approach to correlation functions of resonances in quantum chaotic scattering, {\em Phys. Rev. Lett.} {\bf 83}, 65 (1999).

\bibitem{QCD} J. J. M. Verbaarschot, {\em Handbook Article on 'Applications of Random Matrix Theory to QCD'}, 
invited chapter in “Handbook on Random Matrix Theory”, Eds. G. Akemann, J. Baik, P. Di 
Francesco (2011 Oxford University Press) [arXiv:0910.4134 [hep-th]]. 

\bibitem{Johansson} K. Johansson, Shape fluctuations and random matrices, {\em Comm. Math. Phys.} {\bf 209}, 437 (2000).

\bibitem{MZ1} S. Maslov and Y. C. Zhang, Extracting hidden information from knowledge networks, {\em Phys. Rev. Lett.} {\bf 87}, 248701 (2001).

\bibitem{Z2} Y. K. Yu and Y. C. Zhang, On the Anti-Wishart distribution, {\em Physica A}{\bf 312}, 1 (2002).

\bibitem{JN} R. A. Janik and M. A. Nowak, Wishart and anti-Wishart random matrices, {\em J. Phys. A: Math. Gen.} {\bf 36}, 3629 (2003).

\bibitem{vivolarge}  P. Vivo, S. N. Majumdar, and O. Bohigas, Large deviations of the maximum eigenvalue in Wishart random matrices,
{\em J. Phys. A: Math. Theor.} {\bf 40}, 4317 (2007).

\bibitem{castillo} E. Katzav and I. P. Castillo, Large deviations of the smallest eigenvalue of the Wishart-Laguerre ensemble, {\em Phys. Rev. E} {\bf 82}, 040104 (2010).

\bibitem{nadalmaj} C. Nadal and S. N. Majumdar, Non-intersecting Brownian interfaces and Wishart random matrices, {\em Phys. Rev. E} {\bf 79}, 061117 (2009). 

\bibitem{majreview} S. N. Majumdar,  {\em Handbook Article on 'Extreme eigenvalues of Wishart matrices: application to entangled bipartite system'}, 
invited chapter in “Handbook on Random Matrix Theory”, Eds. G. Akemann, J. Baik, P. Di 
Francesco (2011 Oxford University Press), [arXiv:1005.4515] (2010).

\bibitem{collins} B. Collins, Product of random projections, Jacobi ensembles and universality problems arising from free probability, {\em Probab. Theory Rel. Fields} {\bf 133}, 315 (2005). 

\bibitem{dumitriu} I. Dumitriu and A. Edelman, Matrix models for 
$\beta$-ensembles, {\em J. Math. Phys.} {\bf 43}, 5830 (2002). 

\bibitem{edeljac} A. Edelman and B. D. Sutton, The beta-Jacobi matrix model, the CS decomposition, and generalized singular value problems,
{\em Found. Comput. Math.} {\bf 8}, 259 (2008).

\bibitem{redel} C. E. I. Redelmeier, Genus expansion for real Wishart matrices, {\em Journal of Theoretical Probability}
DOI: 10.1007/s10959-010-0278-7 (2010).

\bibitem{simm} N. Simm and F. Mezzadri, Poster at VI BRUNEL Workshop on Random Matrix Theory (17-18 December 2010); Talk given at 'Random Matrix Theory and Its Applications I',
MSRI - Berkeley (Sep 2010).

\bibitem{kui}  G. Berkolaiko and J. Kuipers, Transport moments beyond the leading order, {\em Preprint} [arXiv:1012.3526] (2010).

\bibitem{corr} J. Ambj\o rn, C. F. Kristjansen, and Y. M. Makeenko, Higher genus correlators for the complex matrix model, {\em Mod. Phys. Lett. A} {\bf 7}, 3187 (1992).

\bibitem{Mehta} M. L. Mehta, {\em Random Matrices}, 3rd Edition
(Elsevier-Academic Press, 2004).

\bibitem{nagao} T. Nagao and M. Wadati, Correlation functions of random matrix ensembles related to classical orthogonal polynomials, {\em J. Phys. Soc. Jpn.} {\bf 60}, 3298 (1997).

\bibitem{ghosh} S. Ghosh and A. Pandey, Skew-orthogonal polynomials and random-matrix ensembles, {\em Phys. Rev. E} {\bf 65}, 046221 (2002).

\bibitem{kumar} S. Kumar and A. Pandey, Crossover ensembles of random matrices and
skew-orthogonal polynomials (unpublished).






\bibitem{vivovivo} P. Vivo and E. Vivo, Transmission eigenvalue densities and moments in chaotic cavities from random matrix theory, {\em J. Phys. A: Math. Theor.} {\bf
41}, 122004 (2008).

\bibitem{novaes2} M. Novaes, Statistics of quantum transport in chaotic cavities with broken time-reversal symmetry, {\em Phys. Rev. B} {\bf 78}, 035337 (2008).

\bibitem{novaes1} M. Novaes, Full counting statistics of chaotic cavities with many open channels, {\em Phys. Rev. B} {\bf 75}, 073304 (2007).







\bibitem{beenakker} C. W. J. Beenakker, Random-matrix theory of quantum transport, {\em Rev. Mod. Phys.} {\bf 69},
731 (1997).

\bibitem{chang} A. M. Chang, H. U. Baranger, L. N. Pfeiffer, and K. W.
West, Weak Localization in chaotic versus nonchaotic cavities: a striking difference in the line shape, {\em Phys. Rev. Lett.} {\bf 73}, 2111 (1994).

\bibitem{marcus} C. M. Marcus, A. J. Rimberg, R. M. Westervelt, P. F. Hopkins, and A. C. Gossard, Conductance fluctuations and chaotic scattering in ballistic microstructures, {\em Phys. Rev. Lett.} {\bf 69}, 506 (1992).

\bibitem{oberholzer} S. Oberholzer, E. V. Sukhorukov, C. Strunk, C.
Sch\"{o}nenberger, T. Heinzel, and M. Holland, Shot noise by quantum scattering in chaotic cavities, {\em Phys. Rev. Lett.}
{\bf 86}, 2114 (2001).

\bibitem{landauer} R. Landauer, Spatial variation of currents and fields due to localized scatterers in metallic conduction, {\em IBM J. Res. Dev.} {\bf 1}, 223
(1957); Electrical resistance of disordered one-dimensional lattices, {\em Phil. Mag.} {\bf 21}, 863 (1970); D. S. Fisher and P. A.
Lee, Relation between conductivity and transmission matrix, {\em Phys. Rev. B} {\bf 23}, 6851 (1981).

\bibitem{buttikerPRL} M. B\"{u}ttiker, Four-terminal phase-coherent conductance, {\em Phys. Rev. Lett.} {\bf 57}, 1761
(1986).

\bibitem{ya} Ya. M. Blanter and M. B\"{u}ttiker, Shot noise in mesoscopic conductors, {\em Phys. Rep.} {\bf
336}, 1 (2000).

\bibitem{muttalib} K. A.
Muttalib, J. L. Pichard, and A. D. Stone, Random-Matrix Theory and universal statistics for disordered quantum conductors, {\em Phys. Rev. Lett.} {\bf 59},
2475 (1987).

\bibitem{stone} A. D. Stone, P. A. Mello, K. A.
Muttalib, and J. L. Pichard, in {\it Mesoscopic Phenomena in
Solids}, edited by B. L. Altshuler, P. A. Lee and R. A. Webb (North
Holland, Amsterdam, 1991).

\bibitem{mellopereira} P. A. Mello, P.
Pereyra, and N. Kumar, Macroscopic approach to multichannel disordered conductors, {\em Annals of Physics} {\bf 181}, 290 (1988).



\bibitem{Dys:new} F. J. Dyson, Statistical theory of the energy levels of complex systems, {\em  J. Math. Phys.} {\bf 3}, 140 (1962); {\bf 3}, 157 (1962); {\bf 3}, 166 (1962).

\bibitem{forrcond} P. J. Forrester, Quantum conductance problems and the Jacobi ensemble, {\em J. Phys. A: Math. Gen.} {\bf
39}, 6861 (2006).

\bibitem{brouwer} P. W. Brouwer and C. W. J. Beenakker, Diagrammatic method of integration over the unitary group, with applications to quantum transport in mesoscopic systems, {\em J. Math. Phys.} {\bf 37}, 4904
(1996).



\bibitem{savin} D. V. Savin and H.-J. Sommers, Shot noise in chaotic cavities with an arbitrary number of open channels, {\em Phys. Rev. B} {\bf
73}, 081307(R) (2006).

\bibitem{savin2}  D. V. Savin, H.-J. Sommers, and W. Wieczorek, Nonlinear statistics of quantum transport in chaotic cavities, {\em Phys. Rev. B} {\bf 77}, 125332 (2008).

\bibitem{luque} J.-G. Luque and P. Vivo, Nonlinear Random Matrix statistics, symmetric functions and hyperdeterminants, {\em J. Phys. A: Math. Theor.} {\bf 43}, 085213 (2010);
C. Carr\'e, M. Deneufch\^atel, J.-G. Luque, and P. Vivo, Asymptotics of Selberg-like integrals: the unitary case and Newton's interpolation formula, {\em J. Math. Phys.} {\bf 51}, 123516 (2010);
M. Novaes, Asymptotics of Selberg-like integrals by lattice path counting, {\em Ann. Phys.} {\bf 326}, 828 (2011).



\bibitem{sommers}  H.-J. Sommers, W. Wieczorek, and D. V. Savin, Statistics of conductance and shot-noise power for chaotic cavities, {\em Acta Phys. Pol. A} {\bf 112}, 691
(2007).

\bibitem{savinnew} B. A. Khoruzhenko, D. V. Savin, and H.-J. Sommers, Systematic approach to statistics of conductance and shot-noise in chaotic cavities, {\em Phys. Rev. B} {\bf 80}, 125301 (2009).

\bibitem{Kanz} V. Al. Osipov and E. Kanzieper, Integrable theory of quantum transport in chaotic cavities, {\em Phys. Rev. Lett.} {\bf 101}, 176804
(2008); Statistics of thermal to shot noise crossover in chaotic cavities, {\em J. Phys. A: Math. Theor.} {\bf 42}, 475101 (2009).

\bibitem{vivoPRL} P. Vivo, S. N. Majumdar, and O. Bohigas, Distributions of conductance and shot noise and associated phase transitions, {\em Phys. Rev. Lett.} {\bf 101}, 216809 (2008); Probability distributions of linear statistics in chaotic cavities and associated phase transitions, {\em Phys. Rev. B} {\bf 81}, 104202 (2010).












\end{thebibliography}
\end{document}